\documentclass[aps,twocolumn,amsmath,amssymb,floatfix,showpacs,
superscriptaddress]{revtex4}

\usepackage[dvips]{graphics}
\usepackage{color}
\definecolor{gold}{rgb}{0.85,0.66,0}
\definecolor{dred}{rgb}{0.3,0.7,0}

\begin{document}

\title{\textcolor{dred}{Magnetic field induced metal-insulator 
transition in a Kagome Nanoribbon}}

\author{Moumita Dey}

\affiliation{Theoretical Condensed Matter Physics Division, Saha
Institute of Nuclear Physics, Sector-I, Block-AF, Bidhannagar,
Kolkata-700 064, India}

\author{Santanu K. Maiti}

\email{santanu.maiti@saha.ac.in}

\affiliation{Theoretical Condensed Matter Physics Division, Saha
Institute of Nuclear Physics, Sector-I, Block-AF, Bidhannagar,
Kolkata-700 064, India}

\affiliation{Department of Physics, Narasinha Dutt College, 129
Belilious Road, Howrah-711 101, India}

\author{S. N. Karmakar}

\affiliation{Theoretical Condensed Matter Physics Division, Saha
Institute of Nuclear Physics, Sector-I, Block-AF, Bidhannagar,
Kolkata-700 064, India}

\begin{abstract}

In the present work we investigate two-terminal electron transport through 
a finite width kagome lattice nanoribbon in presence of a perpendicular 
magnetic field. We employ a simple tight-binding (T-B) Hamiltonian to 
describe the system and obtain the transmission properties by using Green's 
function technique within the framework of Landauer-B\"{u}ttiker formalism. 
After presenting an analytical description of energy dispersion relation 
of a kagome nanoribbon in presence of the magnetic field, we investigate 
numerically the transmittance spectra together with the density of states 
and current-voltage characteristics. It is shown that for a specific value
of the Fermi energy the kagome network can exhibit a magnetic field 
induced metal-insulator transition which is the central investigation 
of this communication. Our analysis may be inspiring in designing 
low-dimensional switching devices. 

\end{abstract}


\maketitle

\section{Introduction}

Remarkable advances in nanotechnology have enabled us to fabricate 
different semiconductor superlattices and optical lattice systems 
which are promising candidates to simulate and investigate a lot of 
rich and exotic quantum phenomena in condensed matter physics e.g., 
Quantum Hall Effect (QHE)~\cite{qhall}, Spin Hall Effect (SHE)~\cite{shall}, 
manifestation of topological insulators~\cite{top1,top2}, etc. Unlike the 
bulk materials, the quantum dot superlattice systems with different geometry 
are easier to design by nanolithography technique and the controllability 
of electron filling in such lattice systems is possible by applying a gate 
voltage~\cite{gate}. Earlier Fukui {\em et al.} have fabricated square, 
triangular and kagome lattices using InAs wires on a GaAs 
substrate~\cite{fukui1,fukui2,fukui3}. In 2001, Abrecht {\em et al.} 
have prepared a two-dimensional square lattice using GaAs and observed the 
Hofstader butterfly in energy spectra by applying a perpendicular magnetic 
field~\cite{abrecht}. Among these various lattice systems kagome lattice 
structure occupies a very special position because of its fascinating 
property of having a ferromagnetic ground state at zero temperature where 
the single particle energy spectra have a complete dispersionless flat band 
in the tight-binding approximation.

In 1993, Mielke and Tasaki~\cite{mielke1,mielke2} have investigated that 
the Coulomb interaction between the degenerate electronic states induces 
ferromagnetism at zero temperature, when the flat band is half-filled with 
electrons. In 2002, Kimura {\em et al.}~\cite{kimura} have shown that the 
flat band is completely destroyed by the application of a perpendicular 
magnetic field, and by calculating the Drude weight (D) in a closed system 
with Hubbard interaction they have predicted that the magnetic field can 
induce a metal-insulator as well as a ferromagnetic-to-paramagnetic 
transition. Ishii and Nakayama in 2004 have shown that the exitonic binding 
energy of a kagome lattice is larger than other two-dimensional ($2$D) or 
even one-dimensional ($1$D) lattices because of the macroscopic degree of 
degeneracy and the localized nature of the flat band states~\cite{ishii1}, 
in contrast to the concept that exitonic binding energy in spatially higher 
dimension is smaller than those of spatially lower dimension. To elucidate 
the transport properties of such unique flat band electronic states, Ishii 
and Nakayama again in 2006, have studied electron transport through a kagome 
\begin{figure}[ht]
{\centering \resizebox*{8.3cm}{3.5cm}
{\includegraphics{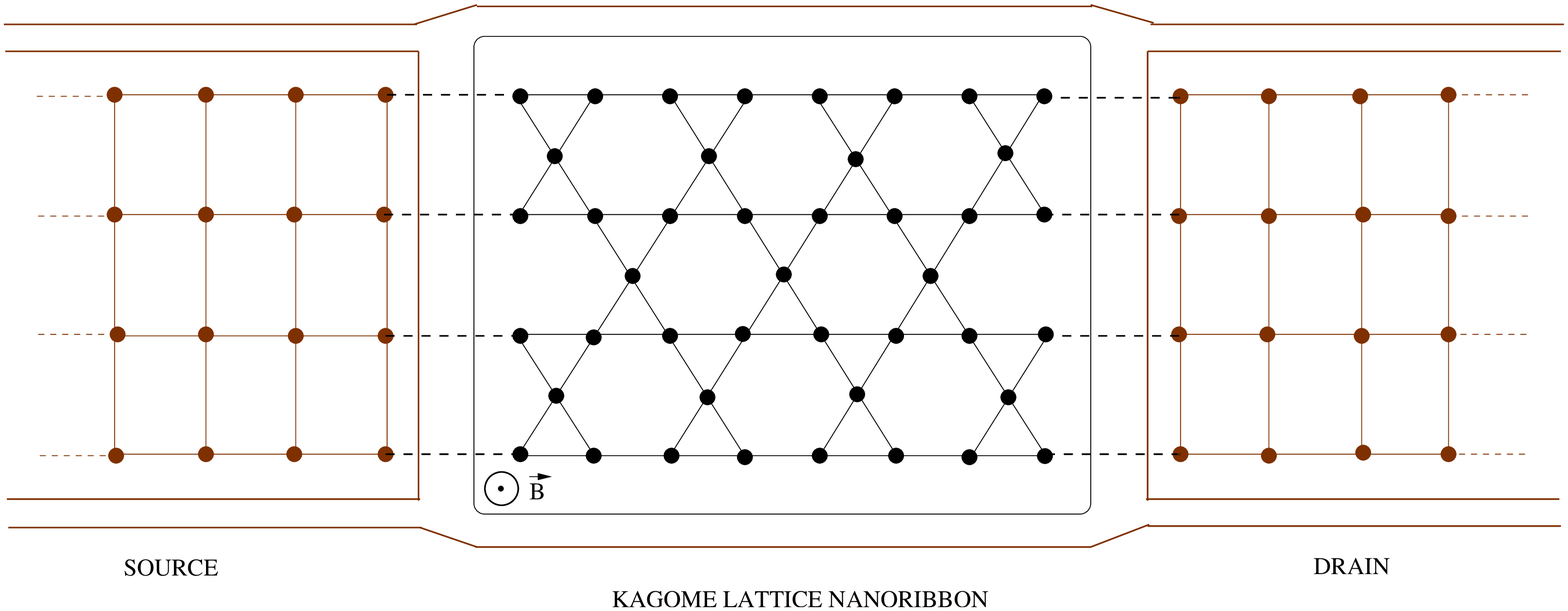}}\par}
\caption{(Color online). Schematic view of a Kagome nanoribbon attached to 
finite width leads, viz, source and drain, in presence of a perpendicular 
magnetic field.}
\label{kagomeribbon}
\end{figure}
lattice chain, in presence of an in-plane electric field~\cite{ishii2}. They 
have observed a large current peak arising from electronic transmission 
through the flat band states when the electric field is applied in 
perpendicular direction to the kagome chain, while no current is observed 
when the field is applied along the chain. These anisotropic features are 
the outcome of the itinerant and localized characteristics of the flat band 
like states which are originated due to quantum interference effect and are 
largely sensitive to the external perturbations like magnetic field or 
electric field.

These are the motivation of our present work where we investigate 
two-terminal quantum transport through a finite size kagome lattice 
ribbon in presence of a perpendicular magnetic field using Green's 
function technique~\cite{datta1,datta2} within Landauer-B{\"u}ttiker 
formalism~\cite{land}. Following an analytical description, in presence
of the magnetic field, of a kagome lattice nanoribbon we study numerically
the transmittance spectra together with the density of states (DOS) and
current-voltage ($I$-$V$) characteristics. Most interestingly we notice 
that the current-voltage characteristics reflect the feature of 
insulator-to-metal transition, when the equilibrium Fermi level $E_F$ is 
fixed at the flat band i.e., $E_F=-2t$ where $t$ being the nearest-neighbor 
hopping strength and the magnetic flux is switched from zero to a non-zero 
one. Again, a similar but reverse transition takes place when the electron 
filling is set at $2t$ i.e., $E_F=2t$. 

The paper is organized as follows. With a brief introduction and 
motivation (Section I), in Section II, we describe the model and 
theoretical formulation to determine the transmission probability, DOS 
and current through the nanostructure. The analytical and numerical 
results are illustrated in Section III. Finally, in Section IV, we 
summarize our results.

\section{Theoretical formulation}

\subsection{Model and Hamiltonian}

We start with Fig~\ref{kagomeribbon}, where a finite width kagome 
lattice ribbon, subject to a perpendicular magnetic field $\vec{B}$, 
is attached to two semi-infinite multi-channel leads, commonly known 
as source and drain. These leads are characterized by the electrochemical 
potentials $\mu_S$ and $\mu_D$, respectively, under the non-equilibrium 
condition when an external bias voltage is applied. Both the leads have 
almost the same cross section as the sample to reduce the effect of the 
scattering induced by wide-to-narrow geometry at the sample-lead interface. 
The whole system is described within a single electron picture by a simple 
tight-binding Hamiltonian with nearest-neighbor hopping approximation.

The Hamiltonian representing the entire system can be written as a 
sum of three terms,
\begin{equation}
H= H_{\small\mbox{kagome}} + H_{\small\mbox{leads}} + H_{\small\mbox{tun}}.
\label{eqn1}
\end{equation}
The first term represents the Hamiltonian for the kagome lattice ribbon
which is coupled to two electron reservoirs through conducting leads i.e.,
source and drain. In Wannier basis the Hamiltonian of the ribbon in
non-interacting picture reads as,
\begin{equation}
H_{\small\mbox{kagome}} = \sum_i \epsilon \, c_{i}^{\dag} c_i + 
\sum_{\langle ij \rangle} \left[\tilde{t}_{ij}\, c_{i}^{\dag}c_{j} + 
h.c.\right]
\label{eqn2}
\end{equation}
where, $\epsilon$ refers to the site energy of an electron at each site of 
the kagome ribbon and $\tilde{t}_{ij}$ corresponds to the nearest-neighbor 
hopping integral between the sites in presence of a perpendicular magnetic 
field. The effect of magnetic field $\vec{B}$ ($=\vec{\nabla}\times\vec{A}$)
is incorporated in the hopping term $\tilde{t}_{ij}$ through the Peierl's 
phase factor and for a chosen gauge field it becomes, 
\begin{equation}
\tilde{t}_{ij} = t\,e^{-\frac{i 2\pi}{\phi_0} \int 
\limits_{\vec{r}_i}^{\vec{r}_j} \vec{A}.\vec{dl}}
\label{eqn3}
\end{equation}
where, $t$ gives the nearest-neighbor hopping integral in the absence 
of magnetic filed and $\phi_0$ ($=ch/e$) is the elementary flux quantum. 
The specific choice of the gauge field in this case and the exact 
\begin{figure}[ht]
{\centering \resizebox*{7.5cm}{6.5cm}{\includegraphics{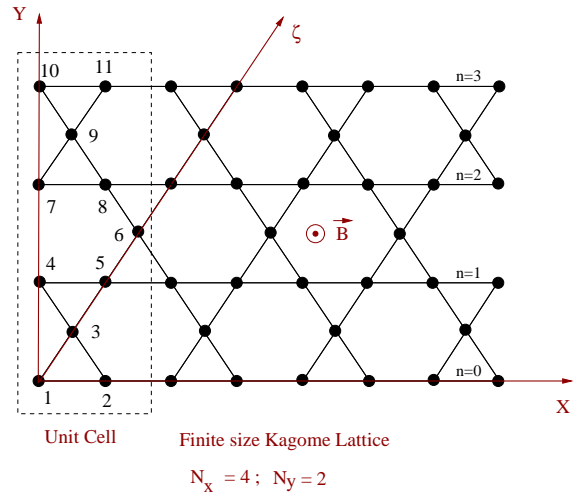}}\par}
\caption{(Color online). Schematic view of a Kagome lattice in presence 
of a perpendicular magnetic field where the unit cell configuration
(dashed region) and the co-ordinate axes are shown.}
\label{kagome1}
\end{figure}
calculation of the Peierl's phase factor has been discussed in the
sub-sequent section. $c_{i}^{\dag}$ and $c_i$ correspond to the creation 
and annihilation operators, respectively, of an electron at the site $i$ 
of the nanoribbon.

The second and third terms of Eq.~\ref{eqn1} describe the T-B Hamiltonians 
for the multi-channel semi-infinite leads and sample-to-lead coupling. In 
Wannier basis they can be written as follows.
\begin{eqnarray}
H_{\small\mbox{leads}} & = & H_S + H_D \nonumber \\
& = & \sum_{\alpha=S,D} \left\{\sum \limits_{n}\epsilon_{l}\,
c_{n}^{\dag} c_n + \sum_{mn} t_l \left[c^{\dag}_{m} c_n + h.c.\right]
\right\}, \nonumber \\
\label{eqn4}
\end{eqnarray}
and,
\begin{equation}
H_{\small\mbox{tun}} = \sum_{S,D} t_c \left[c^{\dag}_{i} c_m + 
c^{\dag}_{m} c_i\right].
\label{eqn5}
\end{equation}
Here, $\epsilon_l$ and $t_l$ stand for the site energy and nearest-neighbor
hopping integral in the leads. $c_{n}^{\dag}$ and $c_{n}$ are the creation 
and annihilation operators, respectively, of an electron at the site $n$ 
of the leads. The hopping integral between the boundary sites of the lead 
and the sample is parametrized by $t_c$. 

\subsection{Calculation of the Peierl's phase factor}

Let us now evaluate $\tilde{t}_{ij}$ incorporating the Peierl's phase 
factor.

We choose the gauge for the vector potential $\vec{A}$ associated 
with the magnetic field $\vec{B}$ ($=B\hat{z}$), perpendicular to the 
lattice plane, in the form,
\begin{equation}
\vec{A} = -By\,\hat{x} + \frac{By}{\sqrt{3}}\,\hat{y} = 
\left(-1,\frac{1}{\sqrt{3}}, 0\right)By.
\label{eqn51}
\end{equation}
This specific choice is followed from a literature~\cite{schr}, and the 
purpose of doing so is solely due to the simplification of the 
factor $\int \vec{A}.\vec{dl}$ along various paths of the ribbon.

With this particular choice of gauge, we determine $\tilde{t}_{ij}$ for 
the three different types of paths of the ribbon through which an electron 
can hop in the following ways.

\vskip 0.25cm
\noindent
$\bullet\,\underline{\textbf{Case\,1:}}$
First, we consider the hopping along the $\zeta$ axis 
(see Fig.~\ref{kagome1}). In this case, our choice of gauge ensures that 
the component of $\vec{A}$ along  $\zeta$ 
\begin{figure}[ht]
{\centering \resizebox*{3.5cm}{4.5cm}{\includegraphics{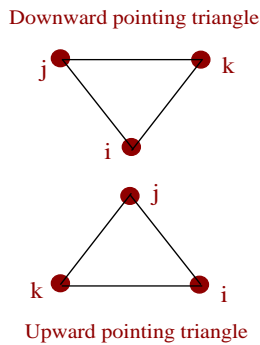}}\par}
\caption{(Color online). Upward and downward pointing triangles labeled
with proper site indices.}
\label{kag}
\end{figure}
axis is zero i.e., $A_{\zeta}=0$. So, $\int \vec{A}.\vec{dl}=0$. Therefore, 
$\tilde{t}_{ij}=t$ when an electron hops along the $\zeta$ axis, either along
the $+ve$, $-ve$ or its parallel direction. 

\vskip 0.25cm
\noindent
$\bullet\,\underline{\textbf{Case\,2:}}$
Here, we consider the hopping along the $X$ ($+$ve for forward hopping and
$-$ve for backward hopping) direction only. In this case 
$\int \vec{A}.\vec{dl} = \int (-By)\,dl = -Bya$. $a$ is the lattice spacing.
Now, for $n=0$ line, $y=0$, and therefore, $\tilde{t}_{ij}=t$. For the line 
$n=1$, $y=\sqrt{3}a$, and hence, $\int \vec{A}.\vec{dl}=-\sqrt{3}Ba^2$. 
If we set $\phi = \frac{\sqrt{3}}{4}Ba^2$, the flux enclosed by the 
smallest triangle (as shown in Fig.~\ref{kag}), then for the line $n=1$, 
$\tilde{t}_{ij}=t\,e^{\frac{i 8 \pi \phi}{\phi_0}}$. Therefore, in general, 
we can write the hopping term for $n$-th line as,
\begin{equation}
\tilde{t}_{ij} = t\,e^{\frac{i 8 \pi n \phi}{\phi_0}}
\label{eqn52}
\end{equation}
where, $n=0,1,2,\ldots \left(2 N_y-1\right)$.

\vskip 0.25cm
\noindent
$\bullet\:\underline{\textbf{Case\:3:}}$
Finally, we consider the case where electrons hop along the direction 
from site $2$ to site $3$ and all its parallel directions. In this case, 
\begin{equation}
\int \limits_{\vec{r}_i}^{\vec{r}_j} \vec{A}.\vec{dl} = \frac{B}{\sqrt{3}} 
\left(y^2_j - y_i^2 \right) = \frac{\sqrt{3}B}{4} \left(\zeta^2_j - 
\zeta_i^2 \right)
\label{equ8}
\end{equation}
It can be shown by straightforward calculation that for an upward pointing 
triangle (shown in Fig.~\ref{kag}) the modified hopping strengths are given 
by,
\begin{equation}
\tilde{t}_{i \rightarrow j} = t\, e^{-\frac{i 8 \pi \phi}{\phi_0}
\left(n+\frac{1}{4}\right)},\hspace{2mm} \mbox{and}, \hspace{2mm}
\tilde{t}_{j \rightarrow i} = t\, e^{\frac{i 8 \pi \phi}{\phi_0}
\left(n+\frac{1}{4}\right)}.
\end{equation}
The value of $n$ belongs to the base line of the triangle.

Similarly, for a downward pointing triangle (shown in Fig.~\ref{kag}) 
the modified hopping integrals can be written as,
\begin{equation}
\tilde{t}_{i \rightarrow j} = t\, e^{-\frac{i 8 \pi \phi}{\phi_0}
\left(n+\frac{3}{4}\right)},\hspace{2mm} \mbox{and}, \hspace{2mm}
\tilde{t}_{j \rightarrow i} = t\, e^{\frac{i 8 \pi \phi}{\phi_0}
\left(n^{\prime}-\frac{1}{4}\right)}.
\end{equation}
Here, $n^{\prime}=n+ 1$, as for a downward pointing triangle the sites 
$i$ and $j$ do not belong to the same value of $n$.

\subsection{Evaluation of the transmission probability and current by 
Green's function technique}

To obtain the transmission probability of an electron through such a
bridge system, we use Green's function formalism. Within the regime of 
coherent transport and in the absence of Coulomb interaction this 
technique is well applied.

The single particle Green's function operator representing the entire
system for an electron with energy $E$ is defined as,
\begin{equation}
G=\left(E - H + i\eta \right)^{-1}
\label{eqn6}
\end{equation}
where, $\eta \rightarrow 0^+$.

Following the matrix form of \mbox{\boldmath $H$} and \mbox{\boldmath $G$}
the problem of finding \mbox{\boldmath $G$} in the full Hilbert space
\mbox{\boldmath $H$} can be mapped exactly to a Green's function
\mbox{\boldmath $G_{\small\mbox{kagome}}^{\small\mbox{eff}}$} 
corresponding to an effective Hamiltonian in the reduced Hilbert space 
of the ribbon itself and we have,
\begin{equation}
\mbox{\boldmath ${\mathcal G}$=$G_{\small\mbox{kagome}}^{\small\mbox{eff}}$}
=\left(\mbox {\boldmath $E- H_{\small\mbox{kagome}} 
-\Sigma_S-\Sigma_D$}\right)^{-1},
\label{equ7}
\end{equation}
where,
\begin{equation}
\mbox{\boldmath $\Sigma_{S(D)}$} = \mbox{\boldmath 
$H_{\small\mbox{tun}}^{\dag} G_{S(D)} H_{\small\mbox{tun}}$}.
\label{eqn8}
\end{equation}
These \mbox{\boldmath $\Sigma_S$} and \mbox{\boldmath $\Sigma_D$} are
the contact self-energies introduced to incorporate the effect of coupling
of the kagome ribbon to the source and drain. It is evident from 
Eq.~\ref{eqn8} that the form of the self-energies are independent of 
the conductor itself through which transmission is studied.

Following Lee and Fisher's expression for the transmission probability 
of an electron from the source to drain we can write,
\begin{equation}
T_{SD} = \mbox{Tr}\mbox{\boldmath [$\Gamma_{\mbox{S}} \mathcal {G}^r 
\Gamma_{\mbox{D}} \mathcal {G}^a$]}.
\label{eqn10}
\end{equation}
$\Gamma_{\alpha}$'s ($\alpha=S$ and $D$) are the coupling matrices 
representing the coupling between the ribbon and the leads and they 
are mathematically defined by the relation,
\begin{equation}
\mbox {\boldmath $\Gamma_{\alpha}$} = i \left[\mbox {\boldmath 
$\Sigma^r_{\alpha} - \Sigma^a_{\alpha}$}\right].
\label{eqn14}
\end{equation}
Here, \mbox{\boldmath $\Sigma_{\alpha}^r$} and 
\mbox{\boldmath $\Sigma_{\alpha}^a$} are the retarded and advanced 
self-energies associated with the $\alpha$-th lead, respectively.

It is shown in literature by Datta {\em et al.}~\cite{datta1,datta2} that 
the self-energy can be expressed as a linear combination of a real and an
imaginary part in the form,
\begin{equation}
\mbox{\boldmath ${\Sigma^r_{\alpha}}$} = \mbox{\boldmath 
$\Lambda_{\alpha}$} - i \mbox{\boldmath $\Delta_{\alpha}$}.
\label{equ17}
\end{equation}
The real part of self-energy describes the shift of the energy levels
and the imaginary part corresponds to the broadening of the levels. The
finite imaginary part appears due to incorporation of the semi-infinite
leads having continuous energy spectrum. Therefore, the coupling matrices
can easily be obtained from the self-energy expression and is expressed as,
\begin{equation}
\mbox{\boldmath $\Gamma_{\alpha}$}=-2\,{\mbox {Im}} 
\left(\mbox{\boldmath $\Sigma_{\alpha}$}\right).
\label{equ18}
\end{equation}
Considering linear transport regime, at absolute zero temperature the
linear conductance $g$ is obtained using two-terminal Landauer
conductance formula,
\begin{equation}
g =\frac{2 e^2}{h}T(E_F).
\label{equ19}
\end{equation}
With the knowledge of the effective transmission probability we compute
the current-voltage ($I$-$V$) characteristics by the standard formalism
based on quantum scattering theory.
\begin{equation}
I(V) = \frac{2 e}{h} \int \limits_{- \infty}^{\infty} 
T \,\left[f_S(E)-f_D(E)\right]\,dE.
\label{eqn20}
\end{equation}
Here, $f_{S(D)}(E) = \left[1+e^{\frac{E-\mu_{S(D)}}{k_{B}T}} \right]^{-1}$
is the Fermi function corresponding to the source and drain. At absolute
zero temperature the above equation boils down to the following expression.
\begin{equation}
I(V)=\frac{2 e}{h} \int \limits_{E_F-\frac{eV}{2}}^{E_F+\frac{eV}{2}} 
T(E) \,dE.
\label{eqn21}
\end{equation}
In our present work we assume that the potential drop takes place only 
at the boundary of the conductor.

\subsection{Evaluation of the self-energy}

Finally, the problem comes to the point of evaluating self-energy for the
finite-width, multi-channel $2$D leads~\cite{verges,nikolic}. Now, for 
the semi-infinite leads (source and drain) as the translational invariance 
is preserved in $X$ direction only, the wave function amplitude at any 
arbitrary site $m$ of the leads can be written as, $\chi_m \propto 
e^{i k_x m_x a} \sin(k_y m_y a)$, with energy 
\begin{equation}
E = 2t_L\left[\cos(k_x a)+ \cos(k_y a)\right].
\label{equ21}
\end{equation}

In Eq.~\ref{equ21}, $k_x$ is continuous, while $k_y$ has discrete values 
given by,
\begin{equation}
k_y(n) = \frac{n \pi}{(m + 1)a}
\label{equ22}
\end{equation}
where, $n= 1,2,3 \ldots m$. $m$ is the total number of transverse 
channels in the leads. In our case $m = 2 N_y$.

The self-energy matrices have non-zero elements only for the sites on the edge
layer of the sample coupled to the leads and it is given by,
\begin{equation}
\Sigma^r_{S(D)}(m,n) = \frac{2}{n_y + 1} \sum_{k_y} \sin(k_y m_y a) 
\Sigma^r (k_y) \sin(k_y n_y a)
\label{equ23}
\end{equation}
where, $\Sigma^{r}(k_y)$ is the self-energy of the each transverse channel 
with a specific value of $k_y$. It is expressed in the following from:
\begin{equation}
\Sigma^r (k_y) = \frac{t_c^2}{2t_L^2} \left[ (E-\epsilon(k_y)) -i 
\sqrt{4t_L^2-(E-\epsilon(k_y))^2} \right]
\end{equation}
with $\epsilon(k_y) = 2t_L \cos(k_y a)$, when the energy lies within 
the band i.e., $|E-\epsilon(k_y)| < 2 t_L$; and,
\begin{equation}
\Sigma^r (k_y) = \frac{t_c^2}{2t_L^2} \left[ (E-\epsilon(k_y)) \mp
\sqrt{(E-\epsilon(k_y))^2-4t_L^2} \right]
\end{equation}
when the energy lies outside the band. The $-ve$ sign comes when 
$E > \epsilon(k_y) + 2|t_L|$, while the $+ve$ sign appears when 
$E < \epsilon(k_y) - 2|t_L|$.

\section{Results and discussion}

We begin by referring the values of different parameters used for our
calculations. Throughout our presentation, we set $\epsilon=\epsilon_l=0$
and fix all the hopping integrals ($t$, $t_l$ and $t_c$) at $-1$. We
measure the energy scale in unit of $t$ and choose the units where
$c=e=h=1$. The magnetic flux $\phi$ is measured in unit of the elementary
flux quantum $\phi_0=ch/e$.

\subsection{Analytical description of energy dispersion relation in 
presence of magnetic field}

To make this present communication a self contained study let us 
start with the energy band structure of a finite width kagome lattice 
nano-ribbon. We obtain it analytically.

Here we follow a general approach to evaluate the band structure of a 
quasi one-dimensional kagome lattice nanoribbon. We establish an effective
difference equation similar to the case of an $1$D infinite chain and this 
can be done by proper choice of a unit cell from the ribbon. The schematic
view of a unit cell is shown by the dashed region of Fig.~\ref{kagome1}.
We consider the ribbon to be made up of the unit cells consisting of 
$N_0$ atomic sites.
\begin{figure}[ht]
{\centering \resizebox*{8.3cm}{13cm}
{\includegraphics{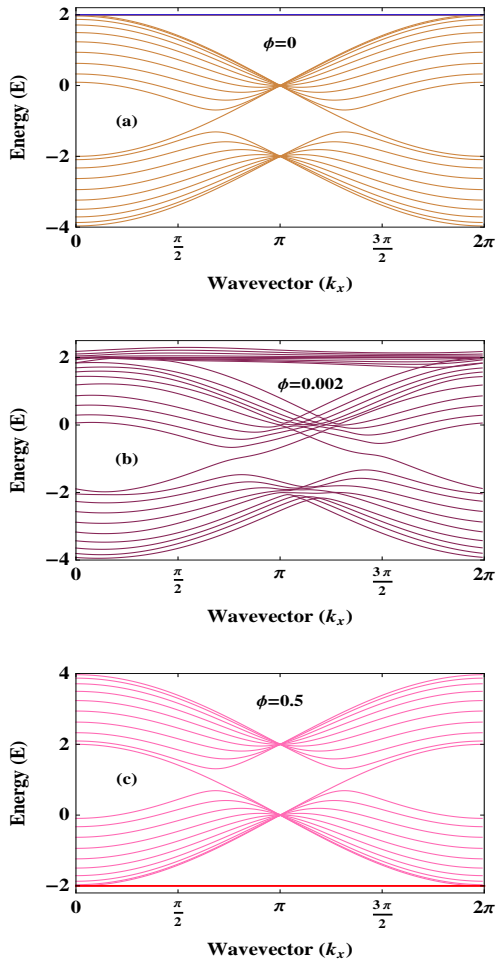}}\par}
\caption{(Color online). Energy levels as a function of $k_x$ for a 
finite width kagome nanoribbon considering $N_y=5$ for different values 
of magnetic flux $\phi$.}
\label{dispersion}
\end{figure}
Within the tight-binding approximation the effective difference equation
for the $n$-th unit cell reads as,  
\begin{equation}
\left(\mbox{\boldmath{$E$}}-\mbox{\boldmath{$\tilde{\epsilon}$}}\right)
\mbox{\boldmath{$\psi_n$}}=\mbox{\boldmath{$\tilde{\tau}\psi_{n+1}$}} +
\mbox{\boldmath{$\tilde{\tau}^{\dag}\psi_{n-1}$}}.
\label{eqn22}
\end{equation}
Here, $\mbox{\boldmath{$\psi_n$}}$ is a column vector with $N_0$ elements
representing the wave function at the $n$-th unit cell. 
$\mbox{\boldmath{$\tilde{\epsilon}$}}$ is a $N_0 \times N_0$ dimensional
matrix and it represents the site-energy matrix of the unit cell.
$\mbox{\boldmath{$\tilde{\tau}$}}$ corresponds to the hopping integral 
between two neighboring unit cells with identical dimension of the 
site-energy matrix. 

Since the system is translationally invariant along the $X$ direction, the
vector $\mbox{\boldmath{$\psi_n$}}$ can be written as,
\begin{equation}
\mbox{\boldmath{$\psi_n$}} = \mbox{\boldmath{$\tilde{A}$}} 
e^{i k_x \lambda_n}.
\label{eqn23}
\end{equation}
Here, 
\begin{equation}
\mbox{\boldmath{$\tilde{A}$}} = \left(\begin{array}{ccc}
A_1 \\
A_2 \\
\vdots \\
A_{N_0}\\
\end{array} \right).
\label{equ24}
\end{equation}
Substituting the value of \mbox{\boldmath{$\psi_n$}} in Eq.~\ref{eqn22} 
we have,
\begin{equation}
\left(\mbox{\boldmath{$E-\tilde{\epsilon}$}}\right) =
\mbox{\boldmath{$\tau$}} e^{i k_x \lambda} + \mbox{\boldmath{$\tau^{\dag}$}} 
e^{-i k_x \lambda}
\label{eqn25}
\end{equation}
The non-trivial solutions of Eq.~\ref{eqn25} are obtained from the 
following condition.
\begin{equation}
\left|\mbox{\boldmath{$E$}} - \mbox{\boldmath{$\tilde{\epsilon}$}} 
-\mbox{\boldmath{$\tau$}} e^{i k_x \lambda} - 
\mbox{\boldmath{$\tau^{\dag}$}} e^{-i k_x \lambda}\right| = 0
\label{eqn26}
\end{equation}
Thus, simplifying Eq.~\ref{eqn26}, we get the desired energy-dispersion
relation ($E$ vs. $k_x$ curve) for the finite width kagome nanoribbon.

As illustrative example, in Fig.~\ref{dispersion} we show the variation
of energy levels as a function of wave vector $k_x$ for a finite width 
\begin{figure}[ht]
{\centering \resizebox*{7.5cm}{5cm}{\includegraphics{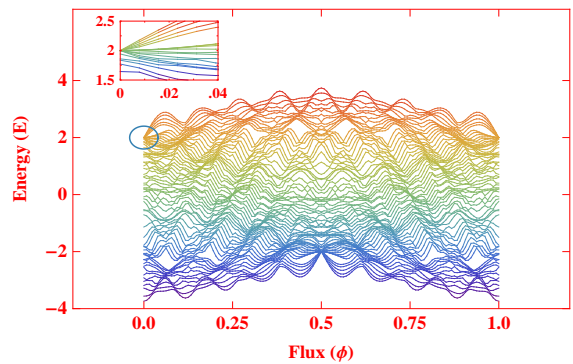}}\par}
\caption{(Color online). Energy levels as a function of flux $\phi$ for a
finite size kagome ribbon considering $N_x=6$ and $N_y=2$. The circular
region is re-plotted in the inset to show the removal of the degeneracy 
with magnetic flux.}
\label{energyflux}
\end{figure}
kagome nanoribbon considering $N_y=5$ for different values of magnetic
flux $\phi$. Most interestingly we see that, when $\phi=0$, a highly
degenerate and dispersionless flat band appears at $E=-2t$ (blue line
of Fig.~\ref{dispersion}(a)) along with the dispersive energy levels.
But, as long as the magnetic flux is switched on, the degeneracy of the
energy levels at the typical energy $E=-2t$ is broken and the flatness
of these levels gets reduced i.e., they start to be dispersive. This
feature is clearly observed from Fig.~\ref{dispersion}(b). Finally, when 
the flux $\phi$ is set at $\phi_0/2$, the flat band again re-appears 
in the spectrum and it situates at the bottom of the energy spectrum
i.e., at $E=2t$ (red line of Fig.~\ref{dispersion}(c)), which is exactly
opposite to that of the case when $\phi=0$. This feature provides the
central idea for the exhibition of magnetic field induced metal-insulator 
transition in a kagome lattice ribbon.
 
\subsection{Energy-flux characteristics}

The behavior of single-particle energy levels as a function of flux $\phi$
for a finite size kagome lattice is shown in Fig.~\ref{energyflux}, where
the energy eigenvalues are obtained by diagonalizing the Hamiltonian
\begin{figure}[ht]
{\centering \resizebox*{8.3cm}{13cm}
{\includegraphics{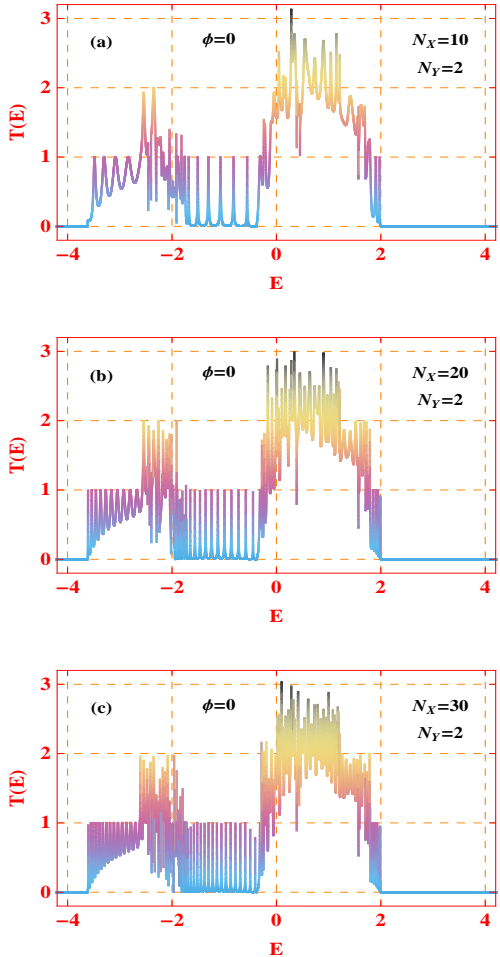}}\par}
\caption{(Color online). Transmission probability as a function of energy
for three different lengths of the kagome nanoribbon when $N_y$ is fixed 
at a particular value ($N_y=2$) in the absence of magnetic flux $\phi$.}
\label{condlength}
\end{figure}
matrix. Here we choose $N_x=6$ and $N_y=2$. This $E$-$\phi$ spectrum is 
sometimes called as Hofstader butterfly spectrum. From the spectrum it
is clearly observed that a highly degenerate energy level is present at
$E=-2t$ when $\phi$ is set at zero, and, the application of a very small
non-zero flux helps to break the degeneracy which is clearly shown in the
inset of Fig.~\ref{energyflux}. The presence of even a very small magnetic 
flux affects the phase of the electronic wave functions and thus destroys 
the quantum interference originated flat band. The flat band again comes 
back because of the quantum interference effect at the energy $E=2t$ when 
$\phi$ is switched to $\phi_0/2$. The $E$-$\phi$ spectrum is periodic in
$\phi$ with periodicity $\phi_0$ ($=1$ in our chosen unit system) and it
is mirror symmetric about $\phi=\phi_0/2$. For an infinitely large
system when $\phi=n/8m$, $n$ and $m$ are two arbitrary integers, $3m$ 
magnetic mini-bands appear in the spectrum, and, for $m=1$ i.e., $\phi=n/8$,
the number of mini-bands is smallest and the gap becomes widest. But in our 
case, we do not observe such gaps, as they are smeared out because of 
the decoherence of the wave functions.

\subsection{Length dependence on transmission probability}

As illustrative example, in Fig.~\ref{condlength}, we show the variation 
of transmission probability as a function of injecting electron energy for
a finite size kagome ribbon in the absence of magnetic flux $\phi$. Here we
fix the width ($N_y=2$) and vary the length $N_x$ of the ribbon. For three
different lengths the results are shown in (a), (b) and (c), respectively.
\begin{figure}[ht]
{\centering \resizebox*{7.5cm}{5cm}{\includegraphics{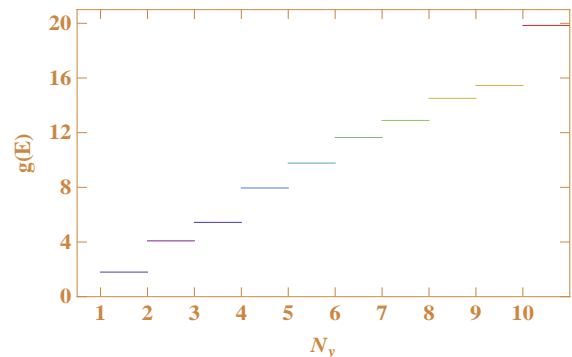}}\par}
\caption{(Color online). Conductance as a function of ribbon width $N_y$ 
for a particular energy $E=0.75$ in the absence of magnetic flux $\phi$.}
\label{condwidth}
\end{figure}
Sharp resonant peaks are observed in the $T$-$E$ spectrum associated with 
the eigenenergies of the ribbon, and therefore, it can be predicted that 
the transmission spectrum manifests itself the electronic structure of the
ribbon. In our numerical calculations, though all the hopping parameters 
are set to the same value ($t=t_L=t_c=-1$), but due to the difference 
\begin{figure*}[ht]
{\centering \resizebox*{8cm}{12.5cm}{\includegraphics{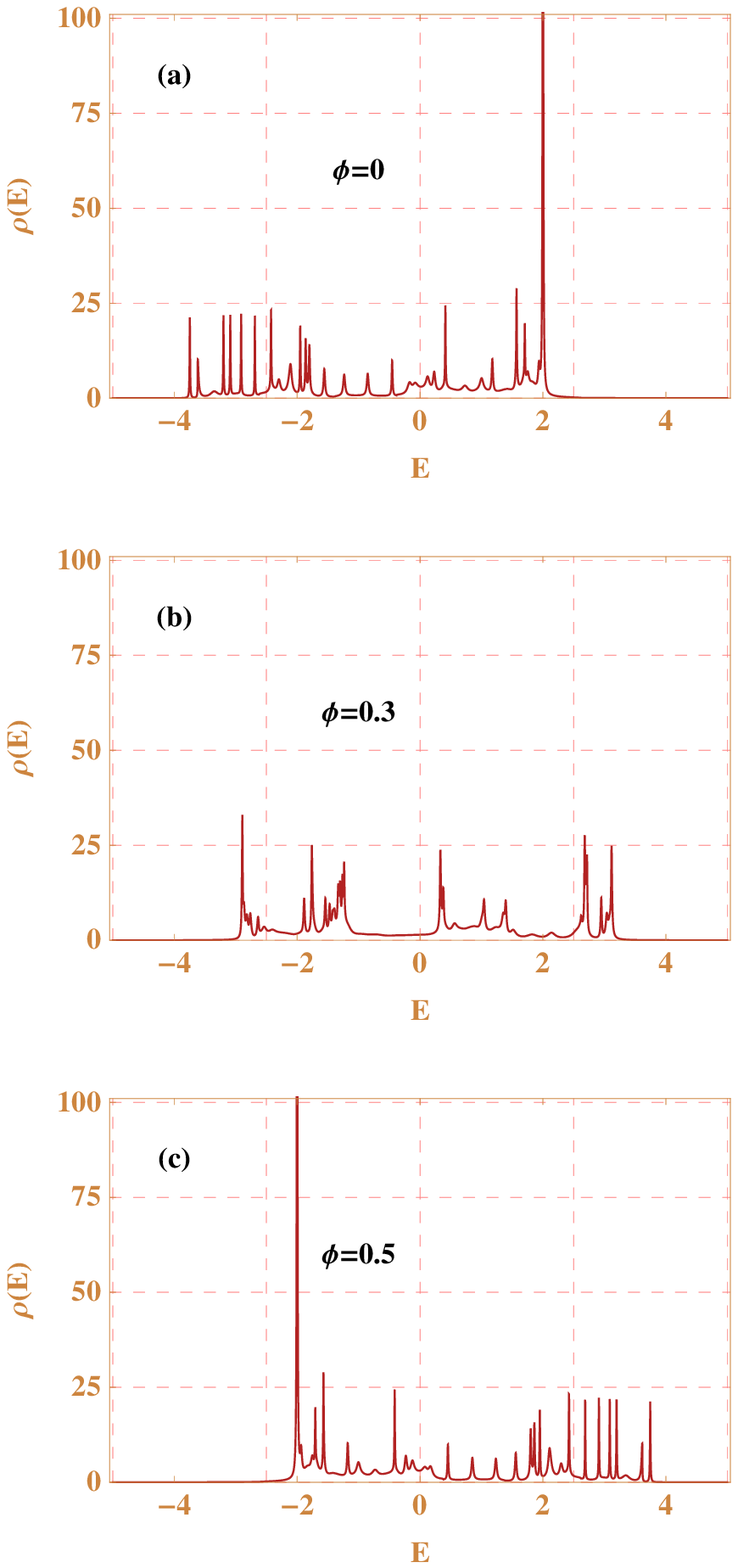}}
        \resizebox*{8cm}{12.5cm}{\includegraphics{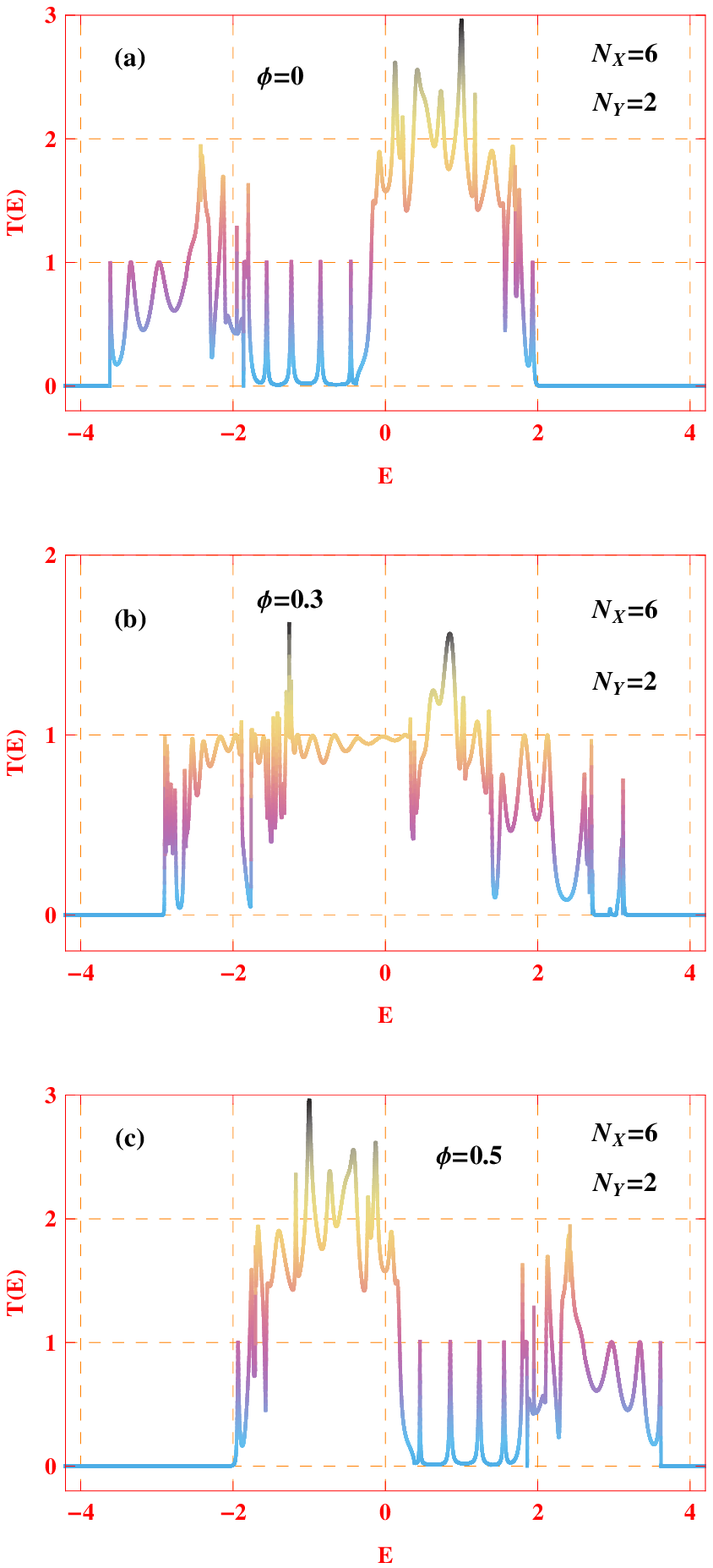}}\par}
\caption{(Color online). Transmission probability (right column) and density 
of states (left column) as a function of energy for a finite size kagome
ribbon ($N_x=6$ and $N_y=2$) for different values of flux $\phi$.}
\label{conddos}
\end{figure*}
in the geometrical structures of the side attached leads (square lattice) 
and conductor (kagome lattice), the geometry has double barrier structure 
which exhibits resonant tunneling conductance. The electrons tunnel from 
the source to drain via the discrete eigenenergy channels of the ribbon. 
The height and width of the transmittance peaks are associated with these
level widths generated through the coupling~\cite{mou1,mou2,san1} of the 
kagome ribbon to the 
leads and it is physically related to the time that an electron stays 
within the ribbon while channeling from the source to drain. As can be 
seen from the figures that the magnitudes of the peaks are not random, 
they are quantized to the integer values which corresponds to the number 
of transverse modes in the leads. With the increase of the length of the 
ribbon, the number of resonant eigenstates are also increased, and 
accordingly, the spectrum looks more dense, but the quantized nature and 
values of $T(E)$ remains unchanged as the ribbon width is kept fixed at a 
certain value.

\subsection{Conductance quantization}

In order to understand the dependence of conductance $g$ on the ribbon
width $N_y$, in Fig.~\ref{condwidth}, we show the variation of two-terminal
conductance as a function of $N_y$ in the absence of magnetic flux when 
$N_x$ is fixed at $30$. The conductance is directly proportional to the
number of transverse modes in the leads which is defined by the factor
$m=2N_y$. Following Landauer formula the conductance at $T=0$\,k is related
to the transmission probability $T({E_F})$ as given in Eq.~\ref{equ19}.
It gives the total transmission probability of an electron through the 
ribbon by adding the net contributions of all the channels. Thus we can 
write,
\begin{equation}
T(E_F) = \sum \limits_{i,j=1}^{m} T_{ij}(E_F)
\end{equation} 
where, $T_{ij}(E_F)$ corresponds to the transmission probability between
the $i$ and $j$-th modes of the two leads. With the increase of $N_y$, the 
number of propagating transverse modes ($m$) also increases, and therefore,
the enhancement in the magnitude of conductance is achieved which is clearly 
shown in Fig.~\ref{condwidth}.

\subsection{Effect of magnetic flux on transport properties}

Now we focus our attention on the effect of magnetic flux on electronic
transport through a finite size kagome ribbon. We illustrate it by 
presenting transmission probabilities and current-voltage characteristics.

\subsubsection{Transmission probability and DOS spectra}

In Fig.~(\ref{conddos}), we present the variation of transmission
probability (right column) along with the nature of density of states 
(DOS) profile (left column) as a function of energy for a finite size 
kagome ribbon considering three different values of magnetic flux $\phi$.
In the top of the left column, DOS is shown when $\phi$ is set at $0$.
A sharp peak is observed at the right edge of the energy band due to 
localized states. These states are highly degenerate and are generally 
pinned at $E=-2t$. The existence of these localized states is a 
characteristic feature of this kind of topology due to quantum 
interference effect between the electronic wave functions. Correspondingly, 
in the transmittance spectrum (top of the right column) we obtain 
transmission peaks at the positions of extended eigenstates but 
{\em no peak} is observed exactly at $E=-2t$ referring to the localized 
states.

The presence of an external magnetic field disturbs the phases of the 
wave functions resulting into annihilation of the localized states and 
\begin{figure}[ht]
{\centering \resizebox*{8cm}{9cm}{\includegraphics{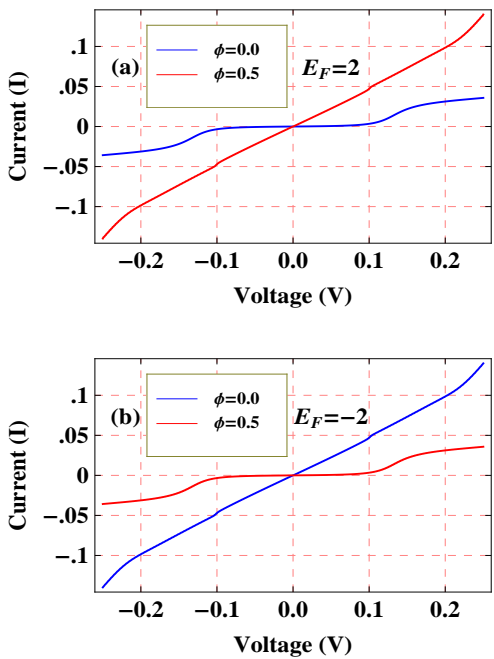}}\par}
\caption{(Color online). Current-voltage characteristics for a finite
size kagome ribbon ($N_x=6$ and $N_y=2$). The possibility of magnetic
field induced metal-insulator transition for two different values of
$E_F$ is shown.}
\label{curr}
\end{figure}
thus a continuum of extended eigenstates is observed as shown in the
middle panel of the left column where $\phi$ is set to an arbitrary value 
$0.3$. Accordingly, the transmission peaks are observed and notably 
finite transmission probability is obtained at $E=-2t$, which was zero 
in the $\phi=0$ case. This feature bears the crucial importance for 
showing the property of insulator-to-metal transition.

Finally, when the magnetic flux is set equal to the half flux quantum 
i.e., $\phi=\phi_0/2$, again the quantum interference effect takes place 
in such a way that the sharp peak associated with the localized states
re-appears in the DOS spectrum (bottom of the left column), but now 
at the left edge ($E=2t$) of the energy band. Similarly, for this energy
transmission probability vanishes and peaks in the transmittance spectrum 
are observed where the extended states are situated (bottom of the right
column).

\subsubsection{Current-voltage characteristics}

Here we explore the possibility of magnetic field induced metal-insulator 
(MI) transition by investigating the current-voltage ($I$-$V$) 
characteristics through a finite size kagome ribbon. It is shown in 
Fig.~\ref{curr}, where two different values of $E_F$ are chosen. The
current through the nano-structure is obtained via Landauer-B\"{u}ttiker
formalism by integrating over the transmission curve (see Eq.~\ref{eqn21}). 
When the equilibrium Fermi level $E_F$ is fixed at $-2t$, almost zero
current is obtained for small bias voltage in the absence of any magnetic
field (blue line in Fig.~\ref{curr}(a)) since the localized states, placed
at $E=-2t$, do not contribute anything in electronic conduction. As the
magnetic field is switched on, localization gets destroyed. Therefore,
setting $E_F$ at the particular value $-2t$, sufficiently large current
is obtained (red line in Fig.~\ref{curr}(a)) compared to the previous 
case where no magnetic field is given which yields the metallic behavior.
From the careful observation we see that when $E_F=-2t$, the best current
magnification is achieved for $\phi=\phi_0/2$.

An exactly similar but reverse transition i.e., metal-to-insulator takes 
place (Fig.~\ref{curr}(b)) when the Fermi level is set at $2t$, instead 
of $-2t$. When $\phi=0$, a large current response (metallic state) is 
observed due to the presence of extended energy eigenstates around $E_F$. 
On the other hand, for $\phi=\phi_0/2$, highly degenerate localized states 
are formed at $E=2t$, and thus, a very low current is achieved in response 
to the applied  bias voltage which leads to the insulating phase. 

\section{Closing remarks}

To summarize, in the present paper we investigate electron transport
properties through a finite size kagome lattice nanoribbon attached to two
finite width leads by using Green's function technique within the framework 
of Landauer-B\"{u}ttiker formalism. The model quantum system is described
by a simple tight-binding Hamiltonian. Following the analytical description
of energy dispersion relation of a finite width kagome nanoribbon in 
presence of magnetic field, we compute numerically the transmittance-energy
spectra together with the density of states and current-voltage 
characteristics. From the energy dispersion curve we see quite interestingly
that, at $\phi=0$, a highly degenerate and dispersionless flat band appears
at $E=-2t$. But as long as the magnetic field is switched on the degeneracy
at this typical energy is broken and the flatness of these levels gets
reduced i.e., they start to be dispersive. Finally, when the flux $\phi$
is set at $\phi_0/2$, the flat band again re-appears and lies at the
bottom of the spectrum i.e., at $E=2t$. This phenomenon provides the 
central idea for the exhibition of magnetic field induced metal-insulator
transition in a kagome lattice nanoribbon and it is justified by studying
the current-voltage characteristics for two different choices of the
equilibrium Fermi energy $E_F$. Our analysis can be utilized in designing
nano-scale switching devices.

The results presented in this communication are worked out for absolute
zero temperature. However, they should remain valid even in a certain
range of finite temperatures ($\sim 300$\,K). This is because the broadening
of the energy levels of the kagome ribbon due to the ribbon-to-lead 
coupling is, in general, much larger than that of the thermal 
broadening~\cite{datta1,datta2}.

\end{document}